\documentclass[epj]{svjour}
\usepackage{graphics}

\begin{document}

\title{Strange and Heavy Flavoured Hypernuclei in
Chiral Soliton Models}

\author{Vladimir B.Kopeliovich \and Andrei M.Shunderuk}

\authorrunning{V.B.Kopeliovich and A.M.Shunderuk}

\institute{Institute for Nuclear Research of Russian Academy of
Sciences, Moscow 117312}

\abstract{The extention of the
chiral soliton approach to hypernuclei - strange or heavy flavoured
 -
becomes more reliable due to success in describing of other properties
of nuclei, e.g. the symmetry energy of nuclei with atomic numbers up
to $\sim 30$. The binding energies of the ground states
of light hypernuclei with $S=-1$ have been described in qualitative agreement
with data. The existence of charmed or beautiful hypernuclei and 
Theta-hypernuclei (strange, charmed or
beautiful) with large binding energy is expected within same approach.
---
}

\maketitle

\section{Main features of the chiral soliton approach. }
The chiral soliton approach (CSA) is based on few
principles and ingredients incorporated in the
truncated {\it effective chiral lagrangian}:
\begin{eqnarray} L^{eff} &=& -{F_\pi^2\over 16}Tr\,(l_\mu l_\mu) +
{1\over 32e^2}Tr [l_\mu l_\nu]^2+ \\ \nonumber
  &+& {F_\pi^2m_\pi^2\over 16}Tr \big(U+U^\dagger -2\big)+...,
\end{eqnarray}
$l_\mu = \partial_\mu U U^\dagger,$ $U\in SU(2)$ or $U\in SU(3)$-
unitary matrix depending on chiral fields, $m_\pi$ is the pion mass,
$F_\pi$-pion decay constant, $e$ - the only parameter of the model.

The soliton (skyrmion) is coherent configuration of classical chiral fields,
possessing topological charge (or winding number) identified with
the baryon number $B$ (Skyrme, 1961).
Important simplifying feature of this approach is that configurations
with different baryon, or atomic numbers are considered on equal footing, when
zero modes only are taken into account in the quantization procedure.
Another feature is that baryons individuality is absent within the
multiskyrmion, and can be recovered - as it is believed - due to careful
consideration of the nonzero modes.

The observed spectrum of baryon states is obtained by means of quantization
procedure and depends on their quantum numbers (isospin, strangeness, etc)
and static characteristics of classical configurations. For the $B=1$
case this was made first in the paper \cite{anw}.
Masses, binding energies of classical configurations with baryon number
$B\geq 2$, their moments of
inertia $\Theta_I,\;\Theta_J$, $\Sigma$-term ($\Gamma$), and some other
characteristics of the chiral solitons contain implicitly
information about interaction between baryons. They are obtained usually
numerically and depend on parameters of the model $F_\pi,\,e$ and masses of
mesons which enter the mass term in the effective lagrangian \footnote{It is of
 interest that baryon interaction potentials depend on 
the weak
 decay constant $F_\pi$ and Skyrme parameter $e$. This connection
of weak and strong interaction properties apparently needs deeper understanding.}.
\section{Ordinary ($S=0$) nuclei; symmetry energy as quantum correction}
In the $SU(2)$ case, which is relevant for description of nonstrange baryons
and nuclei, the rigid rotator quantization model is most adequate when quantum
corrections are not too large.
The symmetry energy $E_{sym}=b_{sym}(N-Z)^2/(2A)$, $b_{sym}\simeq 50\,MeV$,
within chiral soliton approach is described mainly by the isospin dependent quantum
correction
\begin{equation}\delta E_I = {I(I+1)\over 2 \Theta_I},\end{equation}
$\Theta_I \sim A$ being
isotopical moment of inertia, $I=(N-Z)/2$ for the ground states of nuclei.
The $SU(2)$ quantization method - simplest and most reliable - is used here
according to \cite{anw}.
The moment of inertia $\Theta_I$ grows not only with increasing number of
colours, but also with increasing baryon number ($\sim B$ approximately),
therefore this correction decreases like $\sim 1/B$ and such estimates
become more selfconsistent for larger $B$.

\begin{figure}[t]
\begin{center}
\resizebox{0.45\textwidth}{!}{\includegraphics{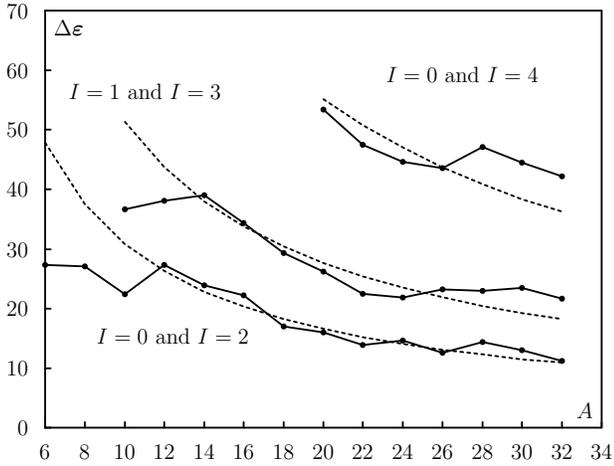}}
\end{center}
\caption{The binding energies differences (in MeV)
for isotopes with even atomic numbers (integer isospins) for the 
"nuclear variant"
of the model with rescaled constant, $e=3.0,\; F_\pi=186 MeV$
(black points connected with solid lines --- experimental data,
dashed lines --- model calculations). The values of isospin of nuclei
which binding energies differences are calculated, are indicated within
the figure.}
\label{fig:1}
\end{figure}

\begin{figure}[t]
\begin{center}
\resizebox{0.45\textwidth}{!}{\includegraphics{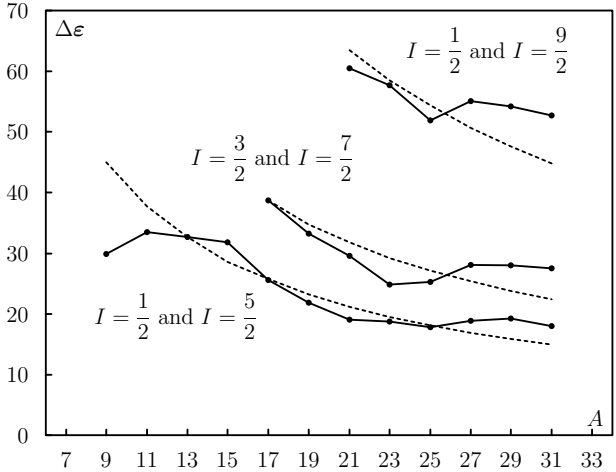}}
\end{center}
\caption{The binding energies differences (in $MeV$) for
isotopes with odd atomic numbers (half-integer isospins) for the rescaled,
or "nuclear" variant of the model,
$e=3.0,\;F_\pi=186 MeV$
. Notations for experimental data and calculation 
results as in Fig.1. Both Fig.1 and Fig.2 are taken from the 
paper \cite{ksm}.}
\label{fig:2}
\end{figure}
In {\bf Fig.}\ref{fig:1} the differences of binding energies between states 
with integer isospins are shown, for the even-even or for the odd-odd nuclei 
(e.g. such differences for the $I=0$ and $I=2$ states or for the $I=1$ 
and $I=3$), calculated 
with the help of formula $(2)$ and for the value of the model parameter $e=3.0$ 
(for the $B=1$ case the value of $e$ is taken usually close to $e\simeq 4.1$ 
which allowed to describe the mass splittings of baryons octet and decuplet). 
Many uncertainties
and some specific corrections introduced in the nuclear mass formula 
are cancelled in such differences (see \cite{ksm} for details and references). 
Similar differences
for the odd-even or even-odd nuclei with half-integer isospins are shown in 
{\bf Fig.}\ref{fig:2}. The differences of binding energies of other nuclei, 
besides those shown in {\bf Fig.}\ref{fig:1} and  {\bf Fig.}\ref{fig:2}, 
are also described well, see \cite{ksm}.
The change of the model parameter $e$ is a natural way to take into account
effectively the nonzero modes - breathing, vibration - which lead to the
increase of dimensions of multiskyrmions (the natural unity of length in the
model is $\sim 1/(F_\pi e)$). Recently similar procedure has
been used for description of the $^6Li$ nucleus \cite{mw} properties.
The change of the pion decay constant $F_\pi$, also made in \cite{mw}, is 
much more
 limited since it is directly measurable quantity, via pion decay.

The mass and baryon number distributions of multiskyrmions have shell-like
form \cite{basu}, at variance from the real ones. However, skyrmions are easily
deformable objects, as previous experience has shown \cite{zks} and recently
has been observed, e.g. for the $B=7$ multiskyrmion \cite{mm}.
 Therefore, one can 
hope that transition to realistic shape of the mass distribution could proceed
without large increase of the energy. Moreover, the important result obtained
recently numerically by Battye, Manton and Sutcliffe \cite{bms} is that at 
large baryon numbers and large enough value of the chiral symmetry 
breaking mass in the lagrangian $(1)$
the transition to more realistic alpha-cluster shape takes place.

The quantum correction due to collective rotation of the multiskyrmion in
usual space equals to $\delta E_J= J(J+1)/(2\Theta_J)$, where $J$ is 
spin of the
 nucleus. It is technically complicated problem to define 
allowed values of $J$ for the
classical configuration with definite symmetry properties. 
The experimentally observed value of spin of the nucleus' ground state
not always can be obtained when
 quantization
of the lowest in energy classical configuration is made \cite{irwin,krusch}. 
However,
due to rich landscape of the classical energy local minima with the energy not 
much
different from the lowest one, but different symmetry properties of the chiral
field configurations\cite{basu}, 
quantization of one of them could give the
 desired value of spin (the $B=7$
case has been considered recently in details in \cite{mm}), and we 
make here in fact a natural assumption that it is always possible.

Since
 the orbital inertia $\Theta_J$ is considerably greater than isotopical 
one
$\Theta_I$ ($\Theta_J \geq B\Theta_I$), this correction is not important
for large enough baryon numbers and is not included in \cite{ksm} and here.
The success of the CSA in description the differences of binding energies
of known nuclei allows to make predictions for the binding energies of
still unknown neutron-rich nuclides (some examples were considered in
\cite{ksm}) and to go further to the consideration of different kinds
of hypernuclei.
\section{Strange hypernuclei ($S=-1$); binding energies of ground states}
\begin{table*}[t]
\caption{The collective motion contributions to the binding energies
of the isoscalar hypernuclei with unit flavour, strangeness or beauty,
$S=-1$ or $b=-1$, in $Mev$. $\Delta \epsilon_{s,b}$, in $Mev$, are
the changes of binding energies of lowest baryonic state with unit flavour,
in comparison with usual nuclei with the same B-number.
$\epsilon^{tot}$ is the total binding energy of the hypernucleus.
Experimental values $\epsilon^{tot}_{exp}$ are taken from \cite{bmz} and
\cite{ht}.
For beauty the first 3 columns correspond to
$r_b=F_B/F_\pi =1.5$, and the last 3 - to $r_b = 2$. }
\label{table:1}
\centering
\begin{tabular}{l l l l l l l l l l l}
\hline
 $_\Lambda A$  &$\omega_s$& $\Delta \epsilon_s $ & $\epsilon^{tot}_s$&
 $\epsilon^{tot}_{exp,s}$
 &$\omega_b^{r_b=1.5}$&$\Delta \epsilon_b$
 &$\epsilon^{tot}_b$
 &$\omega_b^{r_b=2}$&$\Delta \epsilon_b$
 &$\epsilon^{tot}_b$ \\
\hline
$1$ &$306$&---&---&---&$4501$&---&--- &
$4805$&---&--- \\
$^3_\Lambda H$&$ 289$ &$\;-3$  &$\;5$&$2.35$
&$4424$&$75$&$83$
&$4751$&$53$&$61$ \\
$^5_\Lambda He$&$ 287$&$\;-6$& $33 $&$31.4$
&$4422$&$76$&$103$&$4749$&$54$&$81$ \\
$^7_\Lambda Li$&$282 $ &$\;-3$&$29$&$37.2$
&$4429$&$81$&$119$ &$4744$&$59$&$97$ \\
$^9_\Lambda Be$&$291 $&$-13$&$40$&$62.5$
&$4459$&$40$&$97$&$4773$&$31$&$88$ \\
$^{11}_\Lambda B$&$294 $&$-16$&$59$&---
&$4478$&$21$&$96$ &$4786$&$18$&$93$ \\
$^{13}_\Lambda C$&$295$&$-18$&$78$&$104$
&$4488$&$10$&$106$ &$4793$&$11$&$107$ \\
\hline
\end{tabular}
\end{table*}

\begin{table*}[t]
\caption{The binding energies of the isodoublets
of hypernuclei with unit flavour, strangeness or beauty. Other
notations and peculiarities as in {\bf Table 1.} Experimental values
are from \cite{bmz}.}
\label{table:2}
\centering
\begin{tabular}{l l l l l l l l}
\hline
$_\Lambda A$  &$\omega_s$ &$\Delta\epsilon_s$ & $\epsilon^{tot}_s$&
$\epsilon^{tot}_{exp}$
&$\omega_b^{r_b=2}$&$\Delta \epsilon_b$
 &$ \epsilon^{tot}_b$ \\
\hline
$^4_\Lambda H - ^4_\Lambda He$&$283$&$-23$&$5.3$&$10.5;\;\;10.1$&
$4735$&$52$
&$80$ \\
$^6_\Lambda He - ^6_\Lambda Li$&$287$&$-22$&$10.3$&$31.7;\;\;30.8$&
$4752$&$40$
&$72$\\
$^8_\Lambda Li - ^8_\Lambda Be$&$288 $&$-20$&$36.5$&$46.0;\;\;44.4$&
$4765$&$33$
&$89$\\
$^{10}_\Lambda Be -^{10}_\Lambda B$&$292$&$-23$&$42$&$67.3;\;\;65.4$&
$4778$&$20$
&$85$\\
$^{12}_\Lambda B - ^{12}_\Lambda C$&$294$&$-24$&$67$&$87.6;\;\; 84.2$&
$4788$&$11$
&$103$\\
\hline
\end{tabular}
\end{table*}

In the $SU(3)$ case invoking strangeness (or charm, beauty) the flavour
symmetry breaking terms in the lagrangian
\begin{eqnarray}
& &L_{FSB}= \\ \nonumber 
&=&-{F_K^2m_K^2 - F_\pi^2m_\pi^2\over 24}
Tr\left[\left(1-\sqrt 3 \,\lambda_8\right)\left(U+U^\dagger -2\right)\right]+...
\end{eqnarray}
play the crucial role in calculating the spectrum of states with different
flavours (strangeness first of all). Some terms proportional to the difference
$F_K^2-F_\pi^2$ are omitted here (see \cite{kw,vkh,ksh} for details and 
references).
Different
 quantization schemes have been used in literature: rigid rotator, 
soft rotator
 or bound state model.

\begin{figure}[t]

\def\br{\mbox{\boldmath $r$}}
\def\bm{\mbox{\boldmath $m$}}

\setlength{\unitlength}{1.cm}
\hspace*{-40pt}
\begin{picture}(5,6)
\put(3,1.5){\vector(1,0){2.5}}
\put(3,1.5){\vector(0,1){3.8}}
\put(2.8,5.2){$Y$}
\put(5.5,1.2){\bf $I_3$}
\put(1.5,5.7){ a) $Odd \;B\,,\; J=1/2$}
\put(2.3,4.2){$^3 H$}
\put(3.3,4.2){$^3 He$}
\put(2.8,3.3){$^3_\Lambda H$}
\put(2.5,4){\circle* {0.15}}
\put(3.5,4){\circle* {0.15}}
\put(2,3){\circle {0.15}}
\put(3,3){\circle* {0.15}}
\put(3,3){\circle {0.27}}
\put(4,3){\circle {0.15}}
\put(1.5,2){\circle {0.15}}
\put(2.5,2){\circle {0.15}}
\put(3.5,2){\circle {0.15}}
\put(4.5,2){\circle {0.15}}

\end{picture}
\hspace{-15pt}
\begin{picture}(5,6)
\put(3,1.5){\vector(1,0){2.3}}
\put(3,1.5){\vector(0,1){3.8}}
\put(2.8,5.2){$Y$}
\put(5,1.2){$I_3$}
\put(1.5,5.7){ b) $Even \;B\,,\; J=0$}
\put(2.8,4.2){$^4 He $}
\put(2.3,3.3){$^4_\Lambda H$}
\put(3.3,3.3){$^4_\Lambda He$}

\put(3,4){\circle* {0.15}}
\put(2.5,3){\circle*{0.15}}
\put(2.5,3){\circle {0.27}}
\put(3.5,3){\circle*{0.15}}
\put(3.5,3){\circle {0.27}}

\put(2,2){\circle {0.15}}
\put(3,2){\circle {0.15}}
\put(4,2){\circle {0.15}}
\end{picture}
\vspace{-32pt}

\caption{(a) The location of the isoscalar state (shown by double circle)
with odd $B$-number and $|F|=1$ in the upper part of the $(I_3 -Y)$ diagram.
(b) The same for isodoublet states with even $B$. The case of light hypernuclei
$_\Lambda H$ and $_\Lambda He$ is presented as an example. The lower parts of
diagrams with $Y \leq B-3$ are not shown here.}
\label{fig:3}
\end{figure}
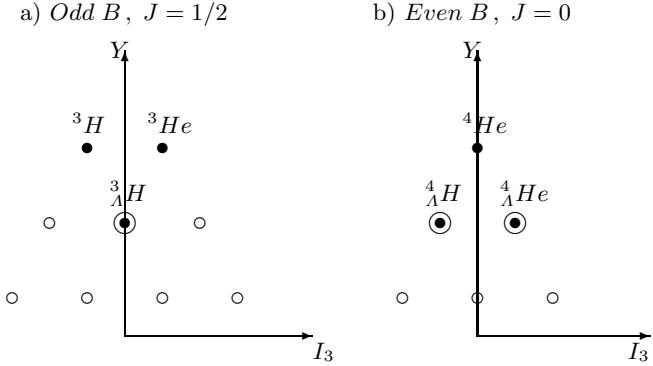

The version of the bound state soliton model proposed by Callan, Klebanov, 
Westerberg
 (1985 - 1996)
and modified for the flavour symmetry breaking case $(F_K > F_\pi)$ allows to
calculate the binding energy differences of ground states between flavoured
and unflavoured nuclei. Combined with few phenomenological arguments it is
very successful in some cases of light hypernuclei.

\begin{figure}[t]
\begin{center}
\resizebox{0.4\textwidth}{!}{\includegraphics{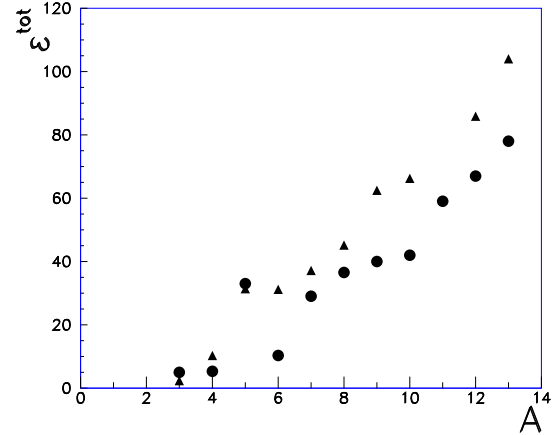}}
\end{center}
\caption{The total binding energies (in MeV) of the ground
states of light $S=-1$ hypernuclei.
Triangles - experimental data, full circles - the collective motion
contribution.}
\label{fig:4}
\end{figure}

Within bound state model (BSM) \cite{kw} (see also \cite{vkh,ksh} for details
and {\bf Fig.3} where location of strange baryon states within minimal $SU(3)$
multiplets is shown)
\begin{equation} M = M_{cl} + \omega_F + \omega_{\bar F} + 
|F| \omega_F + \Delta M_{HFS}\end{equation}
where flavour and antiflavour excitation energies
\begin{equation}\omega_F= N_cB(\mu-1)/8\Theta_F,\;\;\omega_{\bar F}= 
N_cB(\mu+1)/8\Theta_F,
\end{equation}
$\mu \simeq \sqrt{1+\bar m_F^2/M_0^2},\; \bar m_K^2=m_K^2F_K^2/F_\pi^2-m_\pi^2$,
and similar for $D$ or $B$ mesons, 
$M_0^2\simeq N_c^2B^2/(16\Gamma\Theta_F)$, $\Theta_F$ (or $\Theta_K$), 
is the so called
flavour moment of inertia for rotation of the skyrmion to "flavoured direction"
- strange, or charmed, etc., the number of colours $N_c=3$ in all realistic 
calculations. The hyperfine splitting correction $\Delta M_{HFS}$ for the cases
we consider here can be written in the form
\begin{eqnarray}\Delta M_{HFS} &=& {J(J+1)\over 2\Theta_I} + 
(\bar c_F-c_F)\frac{
I_F(I_F+1)}{2\Theta_I}\\ \nonumber
&+&(c_F-1)\frac{[I_r(I_r+1)-I(I+1)]}{2\Theta_I},
\end{eqnarray}
with $I_F=|F|/2=1/2$, and the hyperfine splitting
constants $c_F$ and $\bar c_F$ given by
\begin{equation}c_F =1-{\Theta_I\over 2\mu\Theta_F}(\mu -1);  \qquad
\bar c_F =1-{\Theta_I\over \mu^2\Theta_F}(\mu -1).\end{equation}

There is general qualitative agreement with data in the
behaviour of the calculated binding energy of the ground states of 
$S=-1$ hypernucle1 with increasing atomic number, as can be seen
from Tables 1,2 and {\bf Fig.}4, but the binding
 energy is 
underestimated in most of cases.

The tendency of decrease of binding energies with increasing
$B$-number, beginning with $B\sim 10$, is connected with the fact that
the rational map approximation, leading to the one-shell structure
of the classical configuration, is not good for such values of $B$.

\section{Binding energies of charmed or beautiful hypernuclei}
Same method can be applied for the prediction of the binding energies
of charmed and beautiful hypernuclei \cite{vkh}.
 Evident replacements should 
be made, $m_K\to m_D$ or $m_B$ and $F_K\to F_D$ or $F_B$.

For beautiful hypernuclei the binding energies are presented in Table 1 
(isoscalar states), where the first 3 columns correspond to
$r_b=F_B/F_\pi =1.5$, and the last 3 - to $r_b = 2$, and Table 2 (isodoublets,
$r_b=2$).
\begin{table}[t]
\caption{The binding energies of the charmed hypernuclei, isoscalars
and isodoublets, with unit charm, $c=1$.
$\Delta \epsilon_c$ and $\epsilon^{tot}$ (both in $Mev$) are the same as
in {\bf Tables 1,2}, for the charm quantum number. The results are shown for
two values of charm decay constant $F_D$, corresponding to $r_c=1.5$ and $r_c=2$
(the last $3$ columns). The chemical symbol is ascribed to the nucleus according
to its total electric charge.}
\label{table:3}
\begin{center}
\begin{tabular}{l l l l l l l }
\hline
 $_\Lambda A$ &$\omega_c^{r_c=1.5}$&
$\Delta \epsilon_c$& $\epsilon^{tot}_c$&
 $\omega_c^{r_c=2}$ &
$\Delta \epsilon_c$& $\epsilon^{tot}_c$  \\
\hline
$1$&$1535$&$-$&$-$&$1673$&$-$&$-$ \\
$^3_\Lambda He$&$1504$&$27$&$35$&$1647$&$24$&$32$\\
$^5_\Lambda Li$&$1505$&$25$&$52$&$1646$&$25$&$52$ \\
$^7_\Lambda Be$&$1497$&$32$&$70$&$1641$&$30$&$68$ \\
$^9_\Lambda B$&$1518$&$11$&$68$&$1654$&$17$&$74$\\
$^{11}_\Lambda C$&$1525$&$\;4$&$79$&$1658$&$13$&$87$
\\
$^{13}_\Lambda N$&$1529$&$\;0$&$96$&$1660$&$10$&$106$
\\
\hline
\end{tabular}
\\[12pt]
\begin{tabular}{l l l l l l l}
\hline
 $_\Lambda A$ &$\omega_c^{r_c=1.5}$&$\Delta \epsilon_c$
&$\epsilon^{tot}_c$
 &$\omega_c^{r_c=2}$&$\Delta \epsilon_c$
&$\epsilon^{tot}_c$ \\
\hline
$^4_\Lambda He - ^4_\Lambda Li$    &$1493$&$12$ &$40$&$1639$&$16$ &$44$\\
$^6_\Lambda Li - ^6_\Lambda Be$    &$1504$&$\;9$&$41$&$1646$&$14$ &$46$\\
$^8_\Lambda Be - ^8_\Lambda B$     &$1510$&$\;7$&$63$&$1648$&$15$ &$71$\\
$^{10}_\Lambda B -^{10}_\Lambda C$ &$1520$&$\;0$&$65$&$1655$&$10$ &$75$\\
$^{12}_\Lambda C - ^{12}_\Lambda N$&$1526$&$-4$ &$88$&$1659$&$\;7$&$99$\\
\hline
\end{tabular}
\end{center}
\end{table}

For charmed hypernuclei the binding energies are presented in Table 3 
\cite{vkh}.
Without any new parameters, beautiful (or charmed) hypernuclei are predicted
to be bound stronger than strange hypernuclei. Their binding energies
only slightly depend on the poorly known values of the decay constants $F_D$
or $F_B$ \cite{vkh}.
 There is rough agreement of our results
with some early estimates made within conventional (potential) approach
first by C.Dover and S.Kahana \cite{dk} and later by several authors 
\cite{bb,st}.

The model
we used overestimates the flavour excitation energies, especially for 
strangeness, but is more reliable
for differences of energies which contribute to the differences of binding
energies we calculate here, and for
 charm or beauty quantum numbers
.
The binding energies of states with flavour quantum numbers $|F|=2$ or 
greater have been estimated roughly in \cite{kz}.

\section{Theta-hypernuclei, strange, beautiful or charmed}
Situation with observation of exotic baryons remains to be somewhat
contradictive (see, in particular, the talk by K.Hicks at present conference). 
Apparently, experimental methods of observation of relatively
narrow resonances, with a width about $1 MeV$ or smaller, need
further development. 

Same approach as in previous sections can be applied for the estimates of 
the binding energies of
 Theta- hypernuclei.
 For anti-flavour (positive strangeness, beauty or negative charm) the same 
formula as above holds
, but
with certain changes for the hyperfine splitting constants,
 $c_F \to c_{\bar F}$ and $\bar c_F \to \bar c_{\bar F}$ in the last term 
 $\Delta M_{HFS} $.
$c_{\bar F}$ ($\bar c_{\bar F}$) is obtained from $c_F$ ($\bar c_F$) 
by means of substitution $\mu\to -\mu$:
\begin{equation}c_{\bar F} =1-{\Theta_I\over 2\mu \Theta_F}(\mu +1); \qquad
\bar c_{\bar F} =1+{\Theta_I\over \mu^2\Theta_F}(\mu +1).\end{equation}
This change is crucially important for the link between rotator and
bound state models of the $SU(3)$ quantization \cite{ksh2}, but often it 
was not made in the literature.
The mass of the $\Theta^+$ hyperon within this approach equals to about
$1570\,MeV$ ($e=4.1, \; F_K/F_\pi =1.22$).

\begin{table}[t]
\caption{The collective motion contributions to the binding energies
of the Theta-hypernuclei with unit flavour, strangeness, charm or beauty,
$S=+1$, $c=-1$ and $b=+1$. $\bar \omega_{s,c,b}$, in $Mev$,
 are the antiflavour excitation energies, $\epsilon^{tot}$ is the total 
 binding energy of the
 ground state of
hypernucleus with $|F|=1$. For charm $r_c=1.5$, for beauty $r_b = 2$.}
\label{table:4}
\begin{center}
\begin{tabular}{l l l l l l l}
\hline
 $ A$  &$\bar \omega_s$& $ \epsilon^{tot}_s $ & $\bar \omega_c$&
 $\epsilon^{tot}_c$
 &$\bar \omega_b$ &$\epsilon^{tot}_b$
 \\
\hline
$1$ &$591$ &---  &$1750$&---  &$4940$&---   \\
$3$ &$564$ &$76$ &$1710$&$46$
 &$4890$&$57$  \\
$5$ &$558$ &$108$&$1710$&$71$
&$4880$&$82$  \\
$7$ &$559$ &$120$&$1710$&$85$
&$4880$&$100$\\
$9$ &$550$ &$152$&$1710$&$100$
 &$4900$&$100$\\
$11$&$547$ &$173$&$1710$&$115$&$4900$&$110$\\
$13$&$546$ &$196$&$1720$&$125$&$4910$&$120$\\
\hline
\end{tabular}
\end{center}
\end{table}
As can be seen from Table 4, presenting some of the results obtained in \cite{ksh},
the binding energies for Theta-hypernuclei increase with increasing atomic 
number. If the $\Theta^+$ pentaquark is not narrow and has
the width of several tens of $MeV$, as argued in \cite{ww}, or even greater
,
the Theta-hypernucleus can have much smaller width, and even be bound
relative to the strong interactions \footnote{It should be noted that there is a distinction between different 
quantization schemes in the next-to
leading order of the $1/N_c$ expansion for the flavour symmetry breaking terms 
in the spectrum of baryon states
 \cite{ksh2} - a problem not resolved
yet consistently. The same holds for the widths of baryon resonances, see also
\cite{ww}.}. For rescaled (nuclear) variant of the model with smaller value of the 
parameter $e$, which should be applied for larger atomic numbers, the binding
energies of hypernuclei are greater. In view of these results being in qualitative agreement with
more conventional approaches \cite{miller,cabrera,zhong}, searches for 
such hypernuclear states are of interest.

\section{Conclusions}
Chiral soliton models, based on few principles and
ingredients incorporated in the effective lagrangian, allow to describe
qualitatively, in some cases quantitatively, various chracteristics
of nuclei spectra - from ordinary $(S=0)$ nuclei to known light hypernuclei.
 The symmetry energy of nuclei with isospin up to 4 or 9/2 is described
for atomic numbers between 10 and 30 with only one fixed semifree parameter -
Skyrme constant $e$.

 The binding energies of the ground states of strange hypernuclei have
been described in qualitative and in some cases quantitative agreement with
data for atomic numbers up to $\sim 15$.

In view of this success, predictions of CSM are of interest, including the well
bound heavy 
flavoured (charmed, beautiful)
 hypernuclei and so called Theta-hypernuclei,
i.e. multibaryon states with positive strangeness or beauty, or negative charm.

Essential advantage of this approach is that the case of $B>1$, within CSA, 
does not differ in principle from the case of baryons, until the nonzero modes
are included into consideration.
 There are some
obvious drawbacks of this approach, as continuation of this advantage. 
In particular, one-, two-, etc. baryons
 excitations are not included - this 
is related to a very complicated problem of detailed study
of nonzero modes of multiskyrmions. Specific and  partly technical problem is 
also a smooth
transition from the $B=1$ to rescaled "nuclear variant" of the model with smaller 
value of the parameter $e$.

Some scepticism concerning validity of the CSA - partly because of
the unconfirmed narrow pentaquarks states - has no firm grounds.
Still, the chiral soliton approach is not the complete theory (of course!), 
but may carry
 some important features of the true theory.

The work is supported partly by the RFBR grant 07-02-00960-a.\\

\end{document}